\documentclass[times]{cpeauth}

\usepackage{moreverb}
\usepackage{enumitem}

\newcommand\BibTeX{{\rmfamily B\kern-.05em \textsc{i\kern-.025em b}\kern-.08em
T\kern-.1667em\lower.7ex\hbox{E}\kern-.125emX}}

\def\volumeyear{2012}

\begin{document}


\title{Accelerating R-based Analytics on the Cloud\footnote[2]{This research was fully financially supported by the Natural Sciences and Engineering Research Council of Canada (NSERC) and Flagstone Re, Halifax, Canada under the Collaborative Research and Development grant CRDPJ 412889-11, and was supported in part by an Amazon Web Service Education Grant Program.}}

\author{Ishan Patel, Andrew Rau-Chaplin and Blesson Varghese\corrauth}

\address{\centering Risk Analytics Laboratory, Faculty of Computer Science\\Dalhousie University, Halifax, Nova Scotia, Canada\\E-mail: \{patel, arc, varghese\}@cs.dal.ca}

\corraddr{varghese@cs.dal.ca}

\begin{abstract}
This paper addresses how the benefits of cloud-based infrastructure can be harnessed for analytical workloads. Often the software handling analytical workloads is not developed by a professional programmer, but on an ad hoc basis by Analysts in high-level programming environments such as R or Matlab. The goal of this research is to allow Analysts to take an analytical job that executes on their personal workstations, and with minimum effort execute it on cloud infrastructure and manage both the resources and the data required by the job. If this can be facilitated gracefully, then the Analyst benefits from on-demand resources, low maintenance cost and scalability of computing resources, all of which are offered by the cloud. In this paper, a Platform for Parallel R-based Analytics on the Cloud (P2RAC) that is placed between an Analyst and a cloud infrastructure is proposed and implemented. P2RAC offers a set of command-line tools for managing the resources, such as instances and clusters, the data and the execution of the software on the Amazon Elastic Computing Cloud infrastructure. Experimental studies are pursued using two parallel problems and the results obtained confirm the feasibility of employing P2RAC for solving large-scale analytical problems on the cloud.

\end{abstract}


\maketitle

\vspace{-6pt}

\section{Introduction}
Cloud based infrastructure has proven to be very effective in providing on-demand computational resources to both commercial applications and a wide range of large-scale scientific applications. Applications in climate simulation and analysis \cite{climatesimulationanalysis2}, biomedical image processing \cite{biomedicalimageprocessing1}, satellite data processing \cite{satellitedataprocessing}, astronomy \cite{astronomy1}, and disaster response systems \cite{disasterresponseoperation1} have all been successfully supported using cloud infrastructure. Such successes in both the commercial and scientific settings have depended on bringing to bear the talents of computer scientists and expert developers in order to efficiently exploit cloud-based infrastructure.

In this paper, we explore how a different class of users with a different kind of workload might be able to take advantage of the cloud. In particular, we study how Analysts, who are domain experts with quantitative/mathematical skills, but often with software skills limited to high-level programming environments like R \cite{RProgramming1}, Matlab \cite{matlabandoctave} or Octave \cite{matlabandoctave}, might be supported in harnessing the cloud for ad hoc analytical workloads.

Analytical workloads abound in application domains ranging from computational finance and risk analytics to engineering and manufacturing settings. In our experience, these workloads which often involve simulation \cite{simulation} and optimisation \cite{optimization} tasks share common features as follows:
\begin{enumerate}
\item[(a)] The associated codes are developed by Analysts, not professional developers, in high-level programming environments such as R or Matlab.
\item[(b)] These codes and the related input data are generally created by Analysts for either one time use or are heavily modified each time they are used to adapt them to the analytical question at hand.
\item[(c)] The codes are often computationally intensive or require a large number of independent runs with varying input parameters making some form of parallelism attractive.
\end{enumerate}

Cloud computing is a potential solution that can be beneficial not only to meet the computational requirements of analytical workloads \cite{1} but also to achieve speed-up \cite{2}. A wide range of analytical workloads are already harnessing the benefits of cloud computing. For example, analytical workloads related to data processing \cite{a}, online games \cite{b}, climate \cite{c}, medical records \cite{d}, risk \cite{e}, social networks \cite{f} and neuroscience \cite{h}. 

Our goal has been to develop a platform that allows Analysts to take an analytical job (both the code and associated data) that runs on their personal workstations and with minimum effort and minimum change to the code have them run on large-scale parallel cloud infrastructure. If this can be facilitated gracefully, the Analyst can solve larger problems or perform more experiments in less time. Our approach is somewhat different from other `cluster on cloud' projects such as \cite{project1}\cite{project2}\cite{project3}\cite{project4} in that our focus is to simplify an Analyst's use of cloud infrastructure, rather than provide a fully-configurable high-performance computing cluster on clouds for developers. In particular, we have explored a platform for facilitating R-based risk analytics on the Amazon EC2 cloud. However, we believe that the basic platform and the experienced gained can be generalised to a wider class of analytics and cloud-based infrastructure.

The \textbf{P}latform for \textbf{P}arallel \textbf{R}-based \textbf{A}nalytics on \textbf{C}loud infrastructure (P2RAC) has been designed to harness on-demand compute and storage resources available on the cloud for analytical workloads, and at the same time simplify an Analyst's use of the cloud infrastructure. It provides an interface/API for Analysts to (i) set-up both individual machines and clusters of machines in the cloud, (ii) associate with machines both persistent and short term storage, (iii) transfer both code and data, (iv) execute both batch and interactive jobs, and (v) manage execution and resource termination. The goal is to allow Analysts to submit jobs to the cloud with minimum effort, and manage both the resources and the data required to solve a problem and execute it on the cloud. The current implementation of P2RAC provides support for the resources available on the Amazon AWS infrastructure and offers a set of both core and diagnostic tools, and could also be implemented on other cloud platforms.

In our interaction with Analysts working in industry we have observed a number of commonly arising work patterns. In each of these Analysts start by prototyping an analytical code on their personal workstations, and then as they progress in their work may require additional computing resources. Additional resources are required (i) to perform production runs on a single workstation with more memory or compute speed, (ii) to perform a parameter sweep in which the same code is run hundreds or thousands of times with different input parameters, or (iii) to speed-up a single long running optimisation or simulation task by exploiting co-operative parallelism that may be built into the optimisation or simulation library being used. P2RAC has been designed to address each of these cases. In the later two cases it is important to balance easy-of-use with parallel speed-up. The experimental section of this paper explores this trade-off and shows that reasonable speed-up can often be obtained on cloud infrastructure without undue complexity.

The remainder of this paper is organised as follows. Section \ref{platform} presents the software structure, the sequence of steps that need to be initiated to execute an analytical task on the cloud, and two example workflows for using P2RAC when compute resources, such as instances and clusters are employed. Section \ref{implementation} presents the implementation of P2RAC and the set of commands offered by the platform. Section \ref{experimentalstudies} describes two analytical problems that are executed on the cloud using P2RAC and the results obtained from the experiments. Section \ref{conclusion} concludes the paper.

\section{P2RAC: Platform and Usage}
\label{platform}

\begin{figure}
	\centering
	\includegraphics[width = 0.5\textwidth]{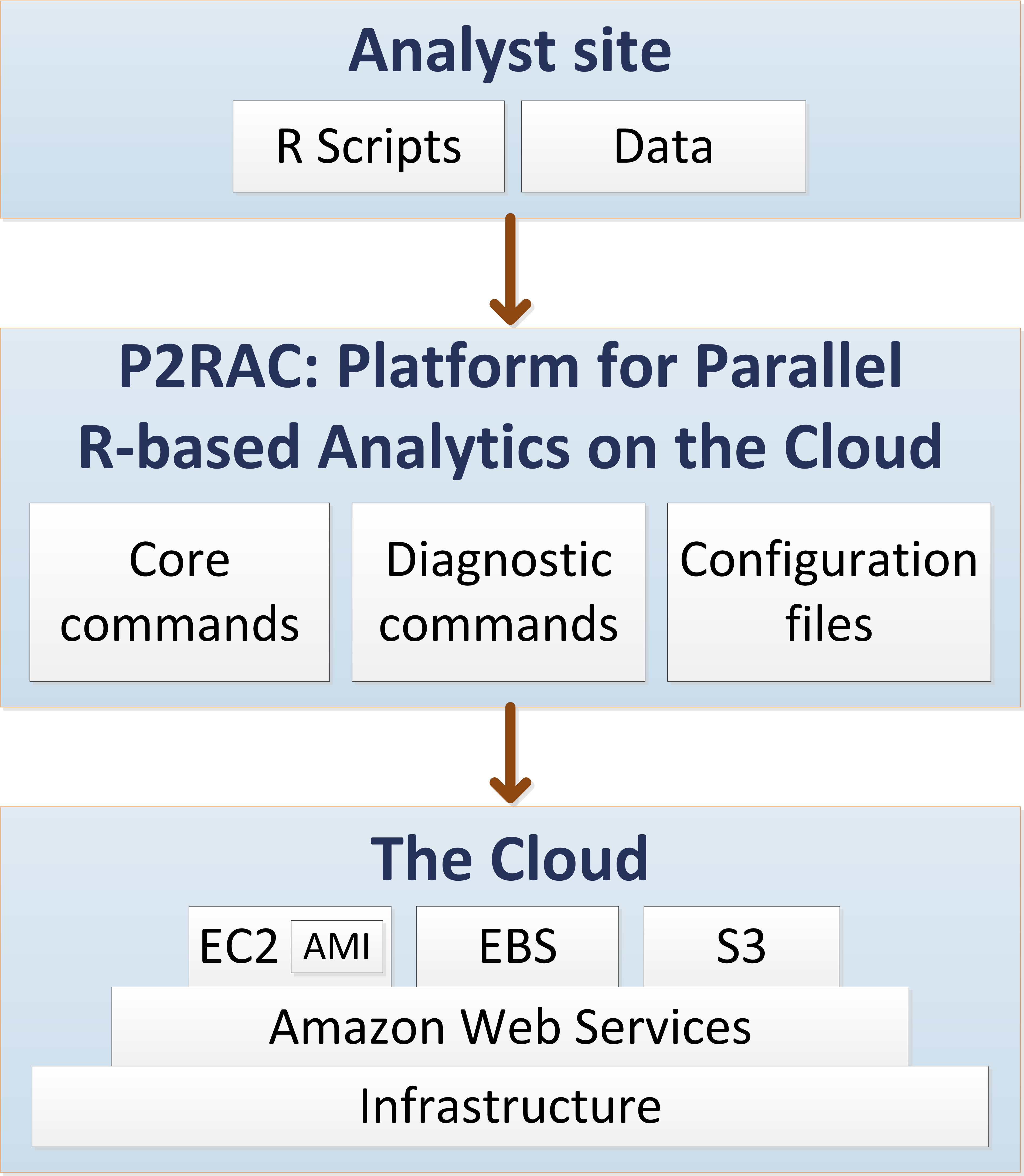}
	\caption{Conceptual design of P2RAC}
	\label{figure1}
\end{figure}

In this section, we discuss the software structure of \textbf{P}latform for \textbf{P}arallel \textbf{R}-based \textbf{A}nalytics on \textbf{C}loud Infrastructure (P2RAC), and its usage. Figure \ref{figure1} illustrates how the platform fits in coherently between an Analyst, who is the typical user, and the cloud infrastructure. The Analyst's project comprising R scripts and large data files is passed on to P2RAC as a task for execution on the cloud. The platform then gathers and initialises a pool of computational and storage resources on the cloud for executing the task. The platform subsequently transfers the task onto the resources and manages task execution. After the task is executed, the platform gathers results which may be spread across the resources in the cloud.

The platform provides support both for an individual computing resource as well as for clusters (a collection of computing resources that work collaboratively). The platform comprises three components, (a) the core tools, (b) the diagnostic tools and (c) the configuration files. The core tools provide functionalities for resource management, data management and execution management. The diagnostic tools provide functionalities for checking the computational environment. The configuration files provide support for the functionalities of the core and diagnostic tools.

A sequence of five steps need to be initiated on an Analyst site to execute a job on the cloud. In the first step, computing and storage resources are initialised on the cloud. In the second step, the analytical project is sent to the resources. In the third step, the scripts within the project are executed on the resources. In the fourth step, results which are generated on the resources are gathered onto the analyst site. In the fifth step, all resources initialised on the cloud are released. 

\subsection{Example Workflows}
\label{workflow}

\begin{figure}
	\centering
	\includegraphics[width = 0.6\textwidth]{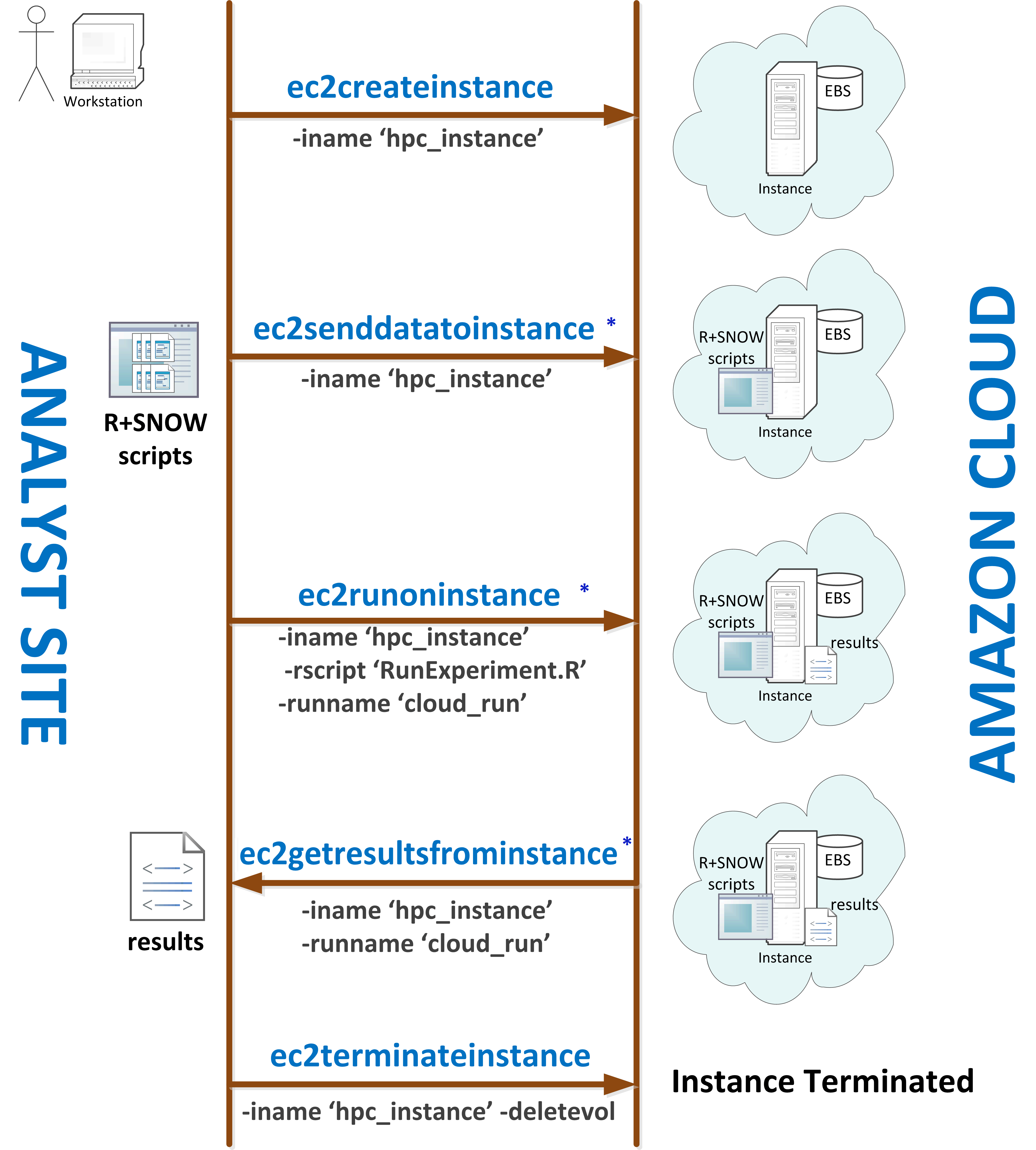}
	\caption{Workflow of commands using P2RAC. * besides a command indicates that the command can be executed multiple times.}
	\label{figure2a}
\end{figure}

The illustration of the above five steps using the command-line tools of P2RAC is shown as two workflows, the first for executing a task on an instance and the second for executing a task on a cluster. For running an R-script on an instance, the sequence of commands are shown in Figure \ref{figure2a}. An Analyst in possession of R scripts utilising SNOW library firstly creates an instance (corresponds to the first step) on the Amazon Cloud. The scripts and the data required by the script are then provided to the instance (corresponds to the second step). The script is executed on the instance and produces the results on the instance (corresponds to the third step). The results can then be fetched by the Analyst (corresponds to the fourth step). Multiple tasks can be executed and their results generated and retrieved from the same instance. Finally, the instance is terminated (corresponds to the fifth step).

\begin{figure}
	\centering
	\includegraphics[width = 0.6\textwidth]{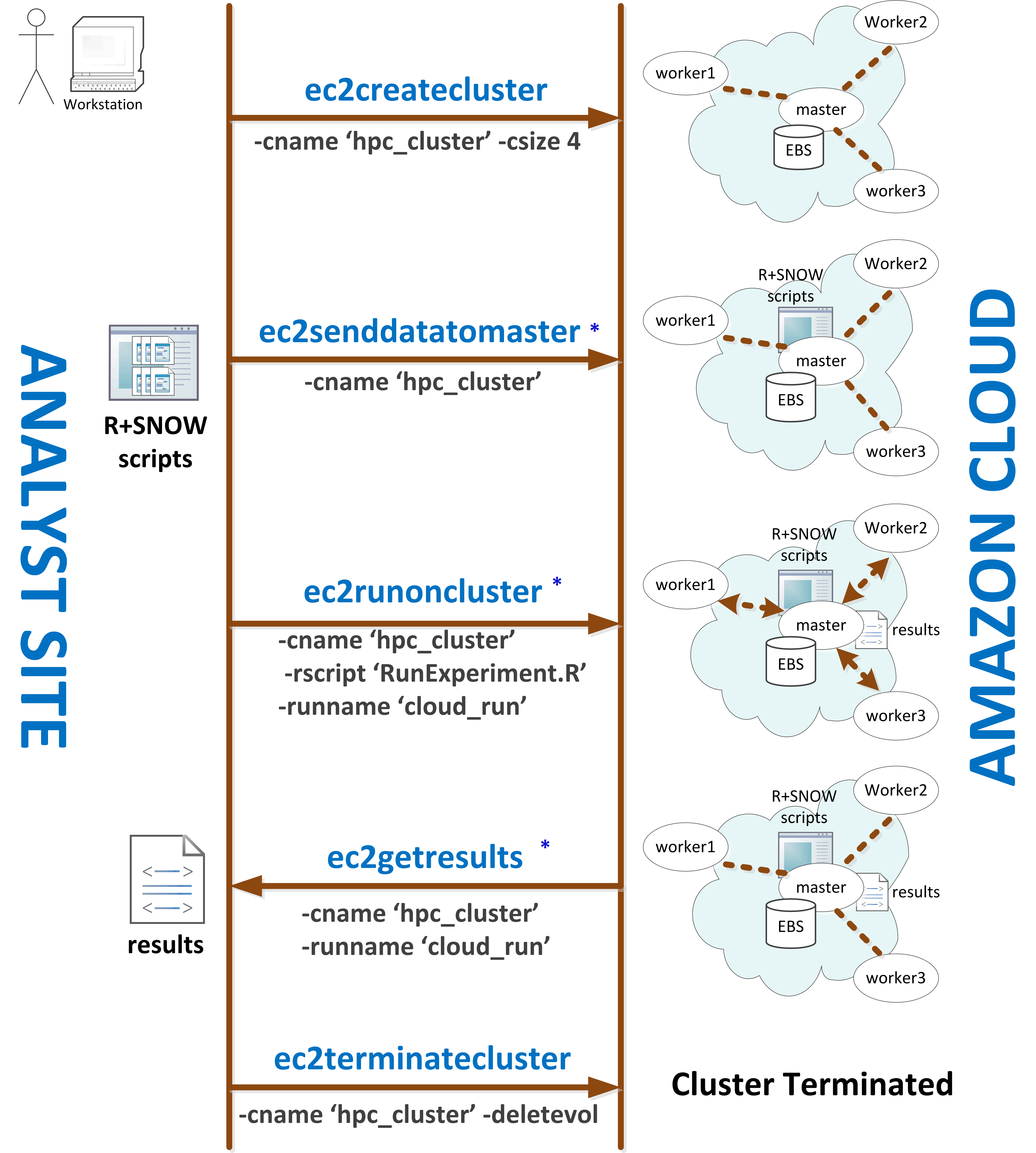}
	\caption{Workflow of commands using P2RAC. * besides a command indicates that the command can be executed multiple times.}
	\label{figure2b}
\end{figure}

Figure \ref{figure2b} shows the sequence of the commands and the order in which they are executed at the Analyst site for executing a task on the cluster. A cluster with 4 instances (1 master and 3 worker instances) is created (corresponds to the first step). The Analyst secondly sends the R scripts and the data required by the scripts to the cluster (corresponds to the second step). The script is then executed and the result of execution is generated on the master instance (corresponds to the third step). When the execution of the R scripts are completed then the Analyst gathers the results to his site (corresponds to the fourth step). Multiple tasks can be executed on the same cluster, and the results generated can be retrieved from the cluster. The cluster is finally terminated (corresponds to the fifth step).

\section{System Design and Implementation}
\label{implementation}

This section considers firstly, how Amazon Web Service (AWS) is supported on P2RAC, and secondly, the tools offered by P2RAC. The tools of P2RAC are separated out as (i) the core tools, (ii) the diagnostic tools and (iii) the configuration files supporting both instances and clusters. This section considers the implementation of the tools that support the cluster and instance. The core and diagnostic tools are implemented as commands which are executed from the command-line. 

\subsection{AWS Support}

The resources provided by a cloud are accessible through Infrastructure as a Service (IaaS) \cite{CloudComputingIaaS}, and Amazon is a leading provider of IaaS. The computational resources on Amazon are referred to as Elastic Compute Cloud (EC2) and are available as instances. These resources are available on-demand and are paid for on the basis of their usage.

The Amazon instances are initialized using Amazon Machine Images (AMI) \cite{AmazonAMIwebsite}. Two Ubuntu AMIs are used in this research. The first AMI, supports Cluster Compute instances offered by Amazon. The Cluster Compute instances are built on Hardware Virtual Machine (HVM) type virtualisation. The second Ubuntu AMI offers support for non-HVM type virtualisation. Any additional libraries which an Analyst needs to include in his project can be installed on the instance by specifying library packages in a configuration file.

The storage resources on Amazon are referred to as the Elastic Block Storage (EBS) \cite{AmazonEBSwebsite}. Similar to EC2, EBS is available on-demand and are paid for on the basis of data transfer and volume of storage. One note worthy feature of EBS includes its ability to provide persistent data storage. This feature is exploited when large volumes of data need to be used in an Analyst's project and thereby eliminates the need for frequent transfer of data that may not be changed over time. Another feature of EBS is that it can be attached (mounted) as a local storage onto an EC2 instance in addition to its storage. This feature eliminates the need for making data locally available to the instance.

The EC2 and EBS services are employed by P2RAC. The P2RAC platform mediates between an Analyst's project and the web services offered by Amazon. The platform is built with the Python programming language and draws heavily on two libraries. Firstly, the BOTO library, which provides P2RAC a Python interface to the Amazon web services. Secondly, the Fabric library, which facilitates remote administration of the instances on the cloud for P2RAC.

\subsection{Core Tools}

The core tools of P2RAC provide functionalities for resource management, data management on the resource, and execution management of a task on the resource. An interface is provided to an Analyst to access the cloud which allows significant computational resources to be brought to bear while greatly reducing the complexities associated with directly working with the cloud infrastructure. This is necessary since an Analyst is less likely to have knowledge and experience of working with computational clouds.

\subsubsection{Instance Support} is offered using tools for instance management, data management on the instance and execution management of a task on the instance.

\textbf{\textit{Instance Management}} tools configure an instance on the cloud, offers the instance for the execution of a task and finally terminates the instance after the task it was executing is completed. P2RAC offers two tools for managing instances, namely \texttt{ec2createinstance} and \texttt{ec2terminateinstance}.

The \texttt{ec2createinstance} command is responsible for configuring an instance on the cloud and making it available to an analyst.

The syntax of the \texttt{ec2createinstance} command is 

\begin{quote}
\texttt{ec2createinstance [-h] [-v] [-iname INSTANCE\_NAME] [-ebsvol EBS\_VOLUME | -snap EBS\_SNAP] [-type INSTANCE\_TYPE] [-desc INSTANCE\_DESCRIPTION]}
\end{quote}

The optional arguments of the \texttt{ec2createinstance} command are \texttt{iname}, \texttt{ebsvol}, \texttt{snap}, \texttt{type} and \texttt{desc}. \texttt{iname} specifies the name of the instance that is created. \texttt{ebsvol} specifies the EBS volume ID which is provided by Amazon when an EBS volume is created. \texttt{snap} specifies the EBS snapshot ID from which an EBS volume can be created. \texttt{snap} and \texttt{ebsvol} cannot be specified at the same time. \texttt{ebsvol} can be specified when a EBS volume is available, while if \texttt{snap} is specified then a new EBS volume is created from the snapshot specified. If both arguments are not provided, then a default snapshot from a configuration file is used. \texttt{type} defines the Amazon EC2 instance type which is specified based on the computational requirements of the task. For example, a High-memory Quadruple Extra Large Instance, offers 68.4 GB of memory, 26 EC2 compute units, 1690 GB of instance storage, with high input/output performance, with instance type as \texttt{m2.4xlarge}, and was employed for a number of experiments in the work reported in this paper. The \texttt{desc} argument can be used to provide a description of the instance.

For example, if a command such as

\begin{quote}
\texttt{ec2createinstance -iname 'hpc\_instance' -ebsvol 'vol-xxxxxxxx' -type 'm2.4xlarge' -desc 'For Trial Simulation Run'}
\end{quote}

\noindent is executed, then a sequence of activities follow. One EC2 instance of \texttt{m2.4xlarge} type is initialised using the AMI specified in the configuration file, and is tagged. The EBS volume, \texttt{vol-xxxxxxxx} (volume ID is masked in this paper) is attached on to the instance. A configuration file at the Analyst site is updated with instance information such as the public DNS names of the instance, EBS volume ID and description of the instance. Should the optional arguments be not provided then the default values which are defined in a configuration file is chosen.

The multiple execution of the \texttt{ec2createinstance} command facilitates the creation of multiple instances. Since an EBS volume can only be attached to one instance, and therefore, the need for multiple EBS volumes arises when multiple instances are created. Should multiple EBS volumes require the same data then they need to snapshot from the same source located on Simple Storage Service (S3) offered by Amazon \cite{AmazonS3website}. Multiple instances cannot have the same name when the \texttt{ec2createinstance} command is executed more than once.

When a task has completed execution on the instance it is essential to safely release the instance. The \texttt{ec2terminateinstance} command is provided. The syntax of the \texttt{ec2terminateinstance} is

\begin{quote}
\texttt{ec2terminateinstance [-h] [-v] [-iname INSTANCE\_NAME] [-deletevol]}\vspace{3pt}
\end{quote}

The optional arguments of the \texttt{ec2terminateinstance} command are \texttt{iname} and \texttt{deletevol}. \texttt{iname} specifies the name of the instance that needs to be terminated. The \texttt{deletevol} switch deletes an EBS volume attached to the instance being terminated. To terminate all the instances on the cloud \texttt{ec2terminateall} command is provided which will be considered in cluster management.

For example, if a command such as 

\begin{quote}
\texttt{ec2terminateinstance –iname 'hpc\_instance'}
\end{quote}

\noindent is executed, then a sequence of activities follows. Firstly, the EBS volume attached to the instance is no more available. Further to this, the instance is terminated. The section containing the instance information of \texttt{hpc\_instance} in the configuration file is removed. Should the \texttt{deletevol} switch be included in the command then the EBS volume, \texttt{vol-xxxxxxxx} is deleted.

\textbf{\textit{Data Management}} on the instance is required to transfer the script and the data from the Analyst's site onto the instance and then receive results. The Secure Copy Protocol (SCP) or the rsync protocol can be employed for data transfer. However, rsync, is chosen as it transfers data quicker than SCP. Moreover, rsync in subsequent data transfers only synchronises the data changed at the source. This feature of rsync makes it suitable for P2RAC since data at the Analyst's site changes frequently. The \texttt{ec2senddatatoinstance} command is provided for data transfer and the results are retrieved using the \texttt{ec2getresultsfrominstance} command.

The syntax of the \texttt{ec2senddatatoinstance} command is 

\begin{quote}
\texttt{ec2senddatatoinstance [-h] [-v] [-iname INSTANCE\_NAME] [-projectdir PROJECT\_DIRECTORY]}
\end{quote}

The optional arguments of the \texttt{ec2senddatatoinstance} command are \texttt{iname} and \texttt{projectdir}. \texttt{iname} specifies the name of the instance to which the project directory will be synchronised. The source project directory is specified as the argument \texttt{projectdir}. If the instance name is not provided by the Analyst then the instance from the configuration file is employed. Should the project directory not be specified then the current working directory at the Analyst site is used as the source project directory. The destination directory is not specified since the project directory is synchronised at the home directory of the root user.

The project directory comprises a set of R scripts which need to be executed, a set of data files required by the scripts and a sub-directory that will contain results after the execution of the script. Large volumes of data which are less likely to change in a short course of time are stored on the EBS volume. On the other hand smaller chunks of data that frequently change are synchronised from an Analyst's site on to the local storage of the instance.

The syntax of the \texttt{ec2getresultsfrominstance} is 

\begin{quote}
\texttt{ec2getresultsfrominstance [-h] [-v] [-iname INSTANCE\_NAME] [-projectdir PROJECT\_DIRECTORY] [-runname RUN\_NAME]}
\end{quote}

The optional arguments are \texttt{iname} and \texttt{projectdir}. The \texttt{iname} argument specifies the name of the instance from where the results have to be fetched. The \texttt{projectdir} specifies the location of the source project directory at the Analyst site. The command utilises the name of the project from the path of the source project directory to fetch data from the corresponding project directory on the instance. If no project directory is specified then the path of the current working directory at the Analyst site is used.

The mandatory argument for the \texttt{ec2getresultsfrominstance} command is \texttt{runname} which indicates the name of a run that was specified during execution and whose results need to be gathered. This argument is used if the same R script has been executed a number of times and each execution had to be differentiated.

\textbf{\textit{Execution Management}} command, namely the \texttt{ec2runoninstance} runs the R script on the instance. This command locks the instance onto the R script and does not permit any additional use of the instance until the script has completed execution or the instance is manually unlocked using \texttt{ec2resourcelock} considered later. The syntax of \texttt{ec2runoninstance} is 

\begin{quote}
\texttt{ec2runoninstance [-h] [-v] [-iname INSTANCE\_NAME] [-projectdir PROJECT\_DIRECTORY] [-rscript R\_SCRIPT] [-runname RUN\_NAME]}
\end{quote}

The optional arguments of \texttt{ec2runoninstance} are \texttt{iname}, \texttt{projectdir} and \texttt{rscript}. The \texttt{iname} argument specifies the name of the instance on which the R script needs to be executed. The \texttt{projectdir} specifies the location of the source project directory at the Analyst site. The command utilises the name of the project from the path of the source project directory to execute an R script from the corresponding project directory on the instance. \texttt{rscript} specifies the name of the R script to be executed from \texttt{projectdir}. If \texttt{rscript} is not provided then the user is prompted to select from a list of R scripts that may be available in the project directory.

The mandatory argument for \texttt{ec2runoninstance} is \texttt{runname} which indicates the name of a run.

\subsubsection{Cluster Support} is offered using tools for cluster management, data management on the cluster and execution management of a task on the cluster.

\textbf{\textit{Cluster Management}} in the core tools are a set of functionalities that range from gathering a pool of instances on the cloud, followed by configuring the instances as a cluster, offering the cluster for task execution and finally terminating the cluster when the task executing on the cluster has completed. P2RAC offers two core tools for cluster management, namely \texttt{ec2createcluster} and \texttt{ec2terminatecluster}.

The \texttt{ec2createcluster} tool is responsible for gathering and configuring the pool of instances as a cluster on the cloud.

The syntax of the \texttt{ec2createcluster} command is 

\begin{quote}
\texttt{ec2createcluster [-h] [-v] [-cname CLUSTER\_NAME] [-csize CLUSTER\_SIZE] [-ebsvol EBS\_VOLUME | -snap EBS\_SNAP] [-type INSTANCE\_TYPE] [-desc CLUSTER\_DESCRIPTION]}
\end{quote}

The optional arguments of the \texttt{ec2createcluster} command are \texttt{cname}, \texttt{csize}, \texttt{ebsvol}, \texttt{snap}, \texttt{type} and \texttt{desc}. \texttt{cname} specifies the name of the cluster that is created. \texttt{csize} specifies the size of the cluster. \texttt{ebsvol} specifies the EBS volume ID which is provided by Amazon when an EBS volume is created. \texttt{snap} specifies the EBS snapshot ID from which an EBS volume can be created. \texttt{snap} and \texttt{ebsvol} cannot be specified at the same time. \texttt{ebsvol} can be specified when a EBS volume is available, while if \texttt{snap} is specified then a new EBS volume is created from the snapshot specified. If both arguments are not provided, then a default snapshot from a configuration file is used. \texttt{type} defines the Amazon EC2 instance type which is specified based on the computational requirements of the task. The \texttt{desc} argument can be used to provide a description of the cluster.

For example, if a command such as 

\begin{quote}
\texttt{ec2createcluster -cname 'hpc\_cluster' -csize '10' -ebsvol 'vol-xxxxxxxx' -type 'm2.4xlarge' -desc 'For Trial Simulation Run'}
\end{quote}

\noindent is executed, then a sequence of activities follow. Ten EC2 instances of \texttt{m2.4xlarge} type are initialized using the AMI specified in the configuration file. The cluster of the ten EC2 instances is referred to as \texttt{hpc\_cluster}. One instance in the \texttt{hpc\_cluster} is denoted as the master and tagged as \texttt{hpc\_cluster\_Master}, while the remaining nine instances are denoted as workers and are tagged as \texttt{hpc\_cluster\_Workers}. The EBS volume, \texttt{vol-xxxxxxxx} (volume ID is masked in this paper) is attached on to the master instance. Network File System (NFS) is employed to share the attached EBS volume among the nine worker instances. A configuration file at the Analyst site is updated with cluster information such as the public DNS names of the master and worker instances, size of the cluster, EBS volume ID, description of the cluster and whether the cluster is in use for executing a script. Should the optional arguments be not provided then the default values which are defined in a configuration file is chosen.

The multiple execution of the \texttt{ec2createcluster} command facilitates the creation of multiple clusters. Since an EBS volume can only be attached to the master instance of one cluster the need for multiple EBS volumes arises when multiple clusters are created. Should multiple EBS volumes require the same data then they need to snapshot from the same source located on Simple Storage Service (S3) offered by Amazon \cite{AmazonS3website}. Multiple clusters cannot have the same name when the \texttt{ec2createcluster} command is executed more than once.

When a task has completed execution it is essential to safely release the resources which are utilised by the cluster. To this end, the \texttt{ec2terminate} cluster command is provided. The syntax of the \texttt{ec2terminatecluster} command is 

\begin{quote}
\texttt{ec2terminatecluster [-h] [-v] [-cname CLUSTER\_NAME] [-deletevol]}
\end{quote}

The optional arguments of the \texttt{ec2terminatecluster} command are \texttt{cname} and \texttt{deletevol}. \texttt{cname} specifies the name of the cluster that needs to be terminated. The \texttt{deletevol} switch deletes an EBS volume attached to the cluster being terminated.

For example, if a command such as 

\begin{quote}
\texttt{ec2terminatecluster –cname 'hpc\_cluster'}
\end{quote}

\noindent is executed, then whether a cluster is in use is firstly checked. If the cluster is in use, then the cluster cannot be terminated. If the cluster is not in use, then a sequence of activities follows. The EBS volume that has been shared with the worker instances through NFS is no more shared. Further to this, the worker instances are terminated such that they do not exist. The EBS volume \texttt{vol-xxxxxxxx} is detached from the master node and the master instance is terminated. The section containing the cluster information of \texttt{hpc\_cluster} in the configuration file is removed. Should the \texttt{deletevol} switch be included in the command then the EBS volume, \texttt{vol-xxxxxxxx} is deleted.

When all resources, both of the instance and the cluster need to be terminated the \texttt{ec2terminateall} command is provided. The syntax of \texttt{ec2terminateall} is

\begin{quote}
\texttt{ec2terminateall [-h] [-v] [-instances] [-clusters] [-ebsvolumes] [-snapshots]}
\end{quote}

The optional switches are \texttt{instances}, \texttt{clusters}, \texttt{ebsvolumes} and \texttt{snapshots} which terminates all the instances, clusters, EBS volumes and snapshots respectively.

\textbf{\textit{Data management}} on the cluster is required to transfer the task (both the scripts and data) from the Analyst site to the cluster on the cloud, and thereafter receive results to the Analyst site. The transfer of task may be to the entire pool of resources in the cluster or to a specific instance on the cluster. Two feasible routes are to use the Secure Copy (SCP) protocol or the rsync protocol. The rsync protocol is employed owing to the quicker synchronisation of data between a source and a destination site. Therefore, two commands, namely the \texttt{ec2senddatatoclusternodes} and \texttt{ec2senddatatomaster} based on the rsync protocol are provided. In order to receive the results to the Analyst site the \texttt{ec2getresults} command is developed.

The \texttt{ec2senddatatoclusternodes} command enables an Analyst's project to be synchronised with all instances of a cluster. This stands different to the data that is stored in an EBS volume mounted on the master instance and shared with the worker instances. In this research, large volumes of data which are less likely to change in a short course of time are stored on the EBS volume. On the other hand smaller chunks of data that frequently change are synchronised from an Analyst's site on to the local storage of the cluster instances.

The structure of a project at the Analyst's site is worthwhile to be noted. A directory comprising a set of R scripts which need to be executed, a set of data files required by the scripts and a sub-directory that will contain results after the execution of the script. The \texttt{ec2senddatatoclusternodes} command synchronises the directory from the Analyst's site to all the instances of the cluster. In other words, every instance contains a project directory.

The syntax of the \texttt{ec2senddatatoclusternodes} command is 

\begin{quote}
\texttt{ec2senddatatoclusternodes [-h] [-v] [-cname CLUSTER\_NAME] [-projectdir PROJECT\_DIRECTORY]}
\end{quote}

The optional arguments of the \texttt{ec2senddatatoclusternodes} command are \texttt{cname} and \texttt{projectdir}. The \texttt{cname} argument specifies the name of the cluster whose instances will be synchronised with the project directory. The source project directory is specified as the argument \texttt{projectdir}. If the cluster name is not provided by the Analyst then the cluster name from the configuration file is employed. Should the project directory not be specified then the current working directory at the Analyst site is used as the source project directory. The destination directory is not specified since the project directory is synchronised at the home directory of the root user.

Owing to the nature of the task to be executed, it may not be necessary that the project directory be provided to all the instances of a cluster. For example, consider a task in which the master instance receives data from the Analyst site and distributes it to the worker instances in the cluster. In such a case it would be inefficient to synchronise the source project directory with all the instances of the cluster, but would be sufficient for the master instance alone to have the project directory. To facilitate this, \texttt{ec2senddatatomaster} command is provided.

The syntax of the \texttt{ec2senddatatomaster} is 

\begin{quote}
\texttt{ec2senddatatomaster [-h] [-v] [–cname CLUSTER\_NAME] [-projectdir PROJECT\_DIRECTORY]}
\end{quote}

The optional arguments of \texttt{ec2senddatatomaster} are similar to that of \texttt{ec2senddatatoclusternodes}.

Based on the nature of the R scripts that are executed there are three possible scenarios for generating results. It is assumed that the R scripts generate results in a sub-directory within the project directory. In the first scenario, the master instance aggregates the results from the worker instances and stores them at the master instance. In the second scenario, however, the results are only generated on the worker instances. In the third scenario, the results are generated on both the master and worker instances. In both the scenarios, the results need to be obtained at the Analyst site. Therefore, the \texttt{ec2getresults} command is provided. To address the first scenario, \texttt{ec2getresults} gathers results from the master instance and provides it at the Analyst site. To address the second scenario, \texttt{ec2getresults} gathers results from the worker instances. In the third scenario, \texttt{ec2getresults} gathers results from both the master and all the worker instances. The aggregated results are stored in a directory at the same hierarchical level of the project directory at the Analyst site.

The syntax of the \texttt{ec2getresults} is 

\begin{quote}
\texttt{ec2getresults [-h] [-v] [–cname CLUSTER\_NAME] [-projectdir PROJECT\_DIRECTORY] [-runname RUN\_NAME] -frommaster | -fromworkers | -fromall}
\end{quote}

The optional arguments are \texttt{cname}, \texttt{projectdir} and a switch. The \texttt{cname} argument specifies the name of the cluster from where the results have to be fetched. The \texttt{projectdir} specifies the location of the source project directory at the Analyst site. The command utilises the name of the project from the path of the source project directory to fetch data from the corresponding project directory on the cluster. If no project directory is specified then the path of the current working directory at the Analyst site is used. The switch specifies the instances from where the results need to be gathered. If \texttt{frommaster} is specified, then results are gathered as in the first scenario. If \texttt{fromworkers} is specified, then results are gathered as in the second scenario. If \texttt{fromall} is specified, then results are gathered as in the third scenario. If no switch is specified then the results are gathered as in the first scenario.

The mandatory argument for the \texttt{ec2getresults} command is \texttt{runname} which indicates the name of a run that was specified during execution and whose results need to be gathered. This argument is used if the same R script has been executed a number of times and each execution had to be differentiated.

\textbf{\textit{Execution Management}} assigns the Analyst task onto a cluster and further runs task on the cluster.  This command locks the cluster for execution of the R script specified by the execution command and does not permit any additional use of the cluster until the script has completed execution or the cluster is manually unlocked using \texttt{ec2resourcelock} considered in the next section. For this, the \texttt{ec2runoncluster} command is provided. The syntax of \texttt{ec2runscript} is

\begin{quote}
\texttt{ec2runoncluster [-h] [-v] [-cname CLUSTER\_NAME] [-projectdir PROJECT\_DIRECTORY] [-rscript R\_SCRIPT] [-runname RUN\_NAME] [-bynode | -byslot]}
\end{quote}

The optional arguments of \texttt{ec2runoncluster} are \texttt{cname}, \texttt{projectdir} and \texttt{rscript}. The \texttt{cname} argument specifies the name of the cluster where the R script needs to be executed. The \texttt{projectdir} specifies the location of the source project directory at the Analyst site. The command utilises the name of the project from the path of the source project directory to execute an R script from the corresponding project directory on the cluster. \texttt{rscript} specifies the name of the R script to be executed from \texttt{projectdir}. If \texttt{rscript} is not provided then the user is prompted to select from a list of R scripts that may be available in the project directory.

Two optional switches to manage the scheduling of slave processes onto the cores of the cluster nodes are available. The \texttt{bynode} switch assigns the processes in a round-robin fashion while the \texttt{byslot} switch assigns all processes on a node until all of its cores are exhausted. In MPI, the default scheduling is \texttt{byslot}, whereas in P2RAC, \texttt{bynode} is chosen as the default scheduling mechanism if the switch is not specified. \texttt{bynode} switch is required to meet the memory constraints of large processes.

The mandatory argument for \texttt{ec2runoncluster} is \texttt{runname} which indicates the name of a run.

\subsection{Diagnostic Tools}
The diagnostic tools support the access of instances and clusters and are available as follows:

\begin{itemize}
\item[(a)] \texttt{ec2listinstances}, \texttt{ec2listclusters} and \texttt{ec2listallresources} - for listing the instances and clusters created by the Analyst on the Amazon cloud
\item[(b)] \texttt{ec2logintoinstance} and \texttt{ec2logintomaster} - for accessing an instance or the master instance of a cluster
\item[(c)] \texttt{ec2resoucelock} - to lock an instance or a cluster for a specific task or to unlock them from use
\end{itemize}

The syntax of \texttt{ec2listinstance} is 

\begin{quote}
\texttt{ec2listinstance [-h] [-v] [-names]}
\end{quote}

The optional switch \texttt{names} provides the names of the instance. If the switch is not provided then the list of the instances, their public DNS names, volume ID of the EBS volume shared with the instances and the description of the cluster is provided.

The syntax of \texttt{ec2listclusters} is 

\begin{quote}
\texttt{ec2listclusters [-h] [-v] [-names]}
\end{quote}

The optional switch \texttt{names} provides the names of the clusters on the cloud. If the switch is not provided then the list of the clusters along with the size of the cluster, public DNS name of all instances, volume ID of the EBS volume shared with the instances of the cluster and the description of the cluster.

The syntax of \texttt{ec2listallresources} is 

\begin{quote}
\texttt{ec2listallresources [-h] [-v] [-instances] [-ebsvols] [-snapshots] [-amis]}
\end{quote}

The switches \texttt{instances}, \texttt{ebsvols}, \texttt{snapshots} and \texttt{amis} provides the names of the instances, EBS volumes, EBS snapshots and AMIs on the cloud.

The syntax of \texttt{ec2logintoinstance} is 

\begin{quote}
\texttt{ec2logintoinstance [-h] [-v] [-iname INSTANCE\_NAME]}
\end{quote}

The optional argument \texttt{iname} specifies the name of the instance that needs to be accessed. The connection to the instance is facilitated through Secure Shell (SSH). If the name of the instance is not provided then the instance listed in the configuration file is used.

The syntax of \texttt{ec2logintocluster} is 

\begin{quote}
\texttt{ec2logintocluster [-h] [-v] [–cname CLUSTER\_NAME]}
\end{quote}

The optional argument \texttt{cname} specifies the name of the cluster whose master instance needs to be accessed. The connection to the master instance is also facilitated through Secure Shell (SSH). If the name of the cluster is not provided then the master instance of the default cluster listed in the configuration file is used.

The syntax of \texttt{ec2resourcelock} is 

\begin{quote}
\texttt{ec2resourcelock [-h] [-v] [-iname INSTANCE\_NAME | –cname CLUSTER\_NAME] [-free | -inuse]}
\end{quote}

The optional arguments \texttt{iname} specifies the name of the instance and \texttt{cname} the name of the cluster that needs to be locked or unlocked. The resources are locked using the \texttt{-inuse} switch and unlocked using the \texttt{-free} switch.

\subsection{Configuration Files}
There are four files that support the core and diagnostic tools which reside on the Analyst site. Firstly, a file that contains a list of variables that are required by the command line tools along with a number of directory paths and references to access keys for Amazon resources. Secondly, a file that provide support for instances, and includes the name of the instance created, its public DNS name, Volume ID, the description of the instance and whether the instance is in use. Thirdly, a file that provides support for clusters, and includes the names of the clusters created, their size, the public DNS names of all their instances, Volume ID of the EBS volume shared by the master with the workers, the description of the clusters and whether the cluster is in use. Fourthly, a file that contains a list of R libraries which are required by an Analyst's project. These libraries are installed on the instances of the cluster when it is created. This is required in addition to the pre-installed libraries of the base AMI.

The P2RAC platform offers support both for instances and clusters. In the case of instance support, P2RAC enables the management of instances, which includes the creation and termination of single and multiple instances. In the case of cluster support, P2RAC facilitates creation and termination of single and multiple clusters. In either case, management of data is facilitated by making large and small chunks of data available to the executing task by sharing persistent data volumes and by synchronising data from the Analyst site to the instances or cluster.

The platform is designed for both batch mode execution and interactive mode execution. Batch-mode execution in P2RAC supports time consuming production tasks. The core commands are listed in a script and the script is executed without the intervention of an Analyst. The interactive mode execution on the other hand allows an Analyst to experiment with his scripts and supports execution of ad hoc tasks. The core commands are executed from the command line by the Analyst. All commands in P2RAC utilise two switches, one of which is \texttt{-h} that provides a description of the use and arguments of the command, and the other which is \texttt{-v} that provides the version of P2RAC.

\section{Experimental Studies}
\label{experimentalstudies}

This section presents the platform, the two problems employed on P2RAC and the results obtained from experiments. The experimental studies is performed for a twofold reason, firstly, to evaluate the feasibility of P2RAC and qualitatively assess its usage and secondly, with respect to parallelism evaluate the speed up that can be expected for both the easy case of independent parallelism and the more challenging case of cooperative parallelism. 


Table \ref{table1} presents the name of the resource and its location, the processor type, the number of cores for the processor, memory, storage capacity and operating system time specifications of the resources. Two desktop computers, two Amazon cloud instances, and four clusters comprising Amazon cloud instances are the computational resources utilised for the evaluation of P2RAC. The Amazon cloud instances are high-memory instances, namely the high-memory double extra large instance (m2.2xlarge) and the high-memory quadruple extra large instance (m2.4xlarge). The four clusters have 2, 4, 8 and 16 nodes (each cluster has one master node) and 8, 16, 32 and 64 available cores (each node has 4 cores) respectively for executing a task. Each m2.2xlarge instance is charged \$0.9 per hour and each m2.4xlarge instance is charged \$1.8 per hour.

\begin{table}
	\begin{center}
	\caption{Resources Utilised for Experimental Studies}
	\begin{tabular}{ l | p{2cm} | p{2.5cm} | p{1cm} | c | c | p{1cm}  }
		\hline
		\hline
		\textbf{Resource}	&	\textbf{Provided by}	&	\textbf{Processor / Amazon Type} 	&	 \textbf{No. of cores}	&	 \textbf{Memory}		&	 \textbf{Storage}	&	\textbf{System type}\\
		\hline
		Desktop A		&	Dalhousie University	&	Intel (R) Core (TM) i7-2600 @ 3.4 GHz	&	8						&	 16GB					&	1.8 TB		& 64 bit\\
		Desktop B		& 	Flagstone Re			&	Intel (R) Xeon X5660 @ 2.8 GHz			&	6				 		&	 24GB					&	2 TB		& 64 bit\\
		Instance A		& 	Amazon					&	m2.2xlarge								&	4				 		&	 34.2 GB				&	850 GB		& 64 bit\\
		Instance B		& 	Amazon					&	m2.4xlarge								&	8				 		&	 68.4 GB				&	1690 GB		& 64 bit\\
		Cluster A		&	Amazon					&	m2.2xlarge X 2							&	8				 		&	 68.4 GB				&	1.7 TB		& 64 bit\\
		Cluster B		&	Amazon					&	m2.2xlarge X 4							&	16				 		&	 136.8 GB				&	3.4 TB		& 64 bit\\
		Cluster C		& 	Amazon					& 	m2.2xlarge X 8							&	32				 		&	 273.6 GB				&	6.8 TB		& 64 bit\\
		Cluster D		& 	Amazon					& 	m2.2xlarge X 16							&	64				 		&	 547.2 GB				& 	13.6 TB		& 64 bit\\	
		\hline \hline
	\end{tabular}
	\label{table1}
	\end{center}
\end{table}


To evaluate the feasibility of P2RAC, two kinds of experimental problems which are analytical in nature were employed. The first problem is a computationally intensive task employing co-operative parallelism, referred to as Catastrophe Bond Optimisation (CATopt). The problem belongs to the domain of reinsurance analytics and is typical of the optimisation problems found in this domain in that (a) it is a large-scale highly non-linear optimisation problem in several thousands of dimensions, (b) it is likely to be executed a few times a year by a sponsoring reinsurance company, and (c) it may require large-scale executions and analysis over the course of several weeks before an actuary can sign-off the results. 

Catastrophe bonds (also known as Cat Bonds) \cite{expt1} are risk-linked securities that transfer a specified set of risks, typically associated with catastrophic loss of property, from a sponsoring insurer or reinsurer to sophisticated investors. If no catastrophe occurs, the sponsor pays a coupon or interest payments to the investors, who made a profitable return. If a catastrophe occurs that meets the conditions described when the bond is issued (referred to as trigger), then the principal would be forgiven and the sponsoring insurance company would use this money to pay claim-holders.

When a cat bond is issued with a parametric trigger, then the investors are made available with (a) regions and perils exposed and the sponsor's share in those region-perils, and (b) the probability of attachment and expected loss. Consider there are $m$ region-peril combinations, for example, Alabama\_Residential or Florida\_Commercial. Then there are $m$ market shares or weights corresponding to the region-perils denoted $w_{j}$, where $j = 1, 2, \ldots , m$. The Industry Losses from an event $i$ can be denoted as $IL_{(i,j)}$, where $j = 1, 2, \ldots , m$, and the sponsor's loss from that event is $sl_{i}$. The loss based on which the sponsor will get his payment is $\sum_{j=1}^{m} w_{j}IL_{(i,j)}$ and the recovery value is $Recovery_{i} = Min(Max(\sum_{j=1}^{m} w_{j}IL_{(i,j)} - Att, 0), Limit)$, where $Att$ is an attachment point which is a deductible and $Limit$ is the maximum payout defined contractually. So the sponsor faces `basis risk' since the actual loss $cl_{i}$ could be very different from $\sum_{j=1}^{m} w_{j}IL_{(i,j)}$, and consequently receive more/less recovery than required. 

In the experimental studies, an R-based example of the CATopt problem which seeks to identify a set of weights that minimizes basis risk is employed. The dimension of the optimisation space is typically 2000-4000 region-perils combinations and there are a number of non-linear constraints that must be applied thereby making the CATopt problem a challenging and computationally intensive problem. The CATopt R script is structured as a distributed genetic algorithm using the rgenoud \cite{rgenoud} R package which combines evolutionary search algorithms with derivative-based (Newton or quasi-Newton) methods. The input data to the CATopt problem is approximately 300 MB.

The second problem is a parameter sweep task \cite{expt2, expt2a} that runs multiple independent jobs without any data dependency between the runs. An R-based example of a simple Monte Carlo simulation was employed. The input data to the parameter sweep task is approximately only 3 MB.

\begin{figure}
	\centering
	\includegraphics[width = \textwidth]{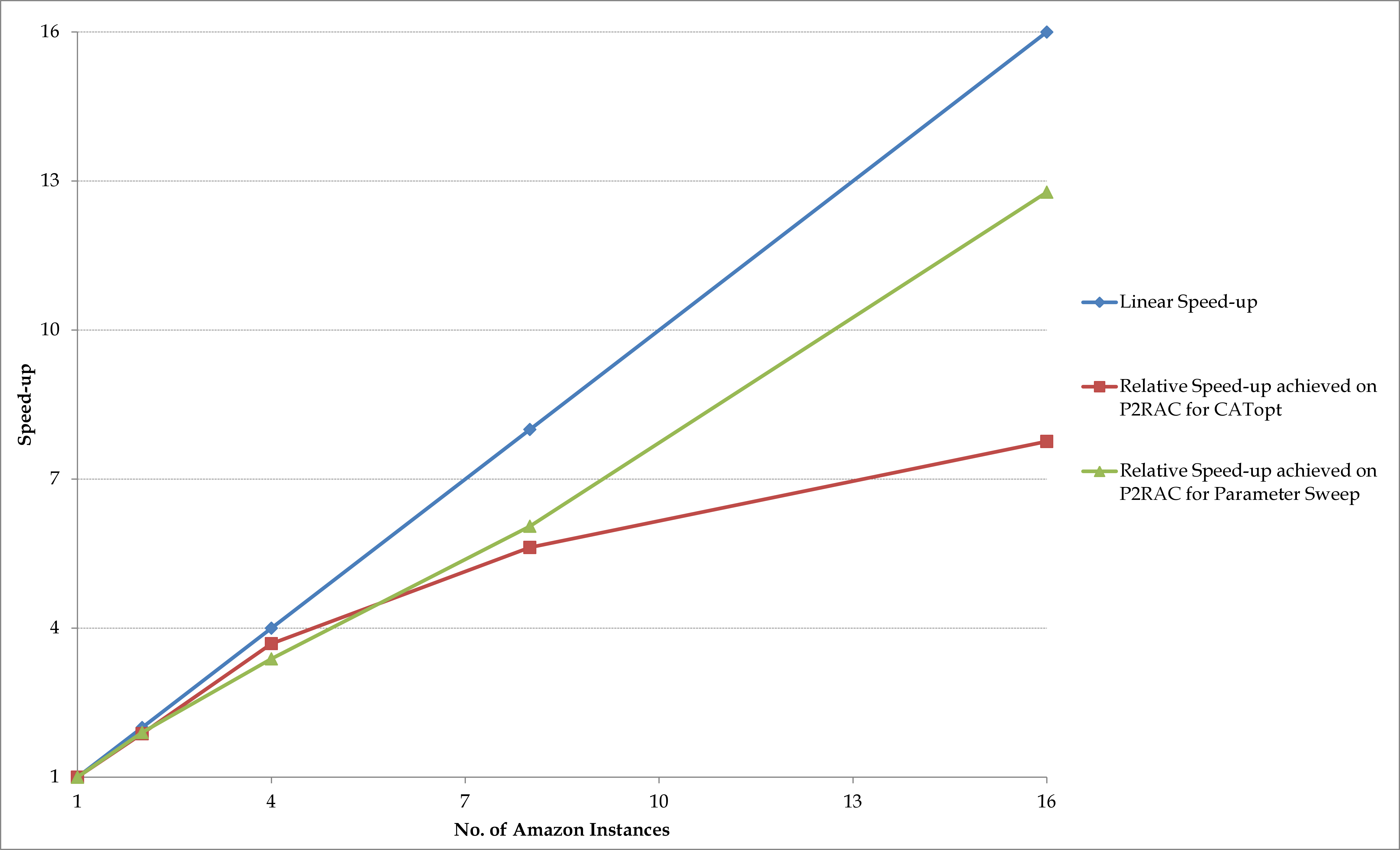}
	\caption{Speed-up achieved for the CATopt and Parameter Sweep Problems using P2RAC}
	\label{graph1}
\end{figure}
\begin{figure}
	\centering
	\includegraphics[width = \textwidth]{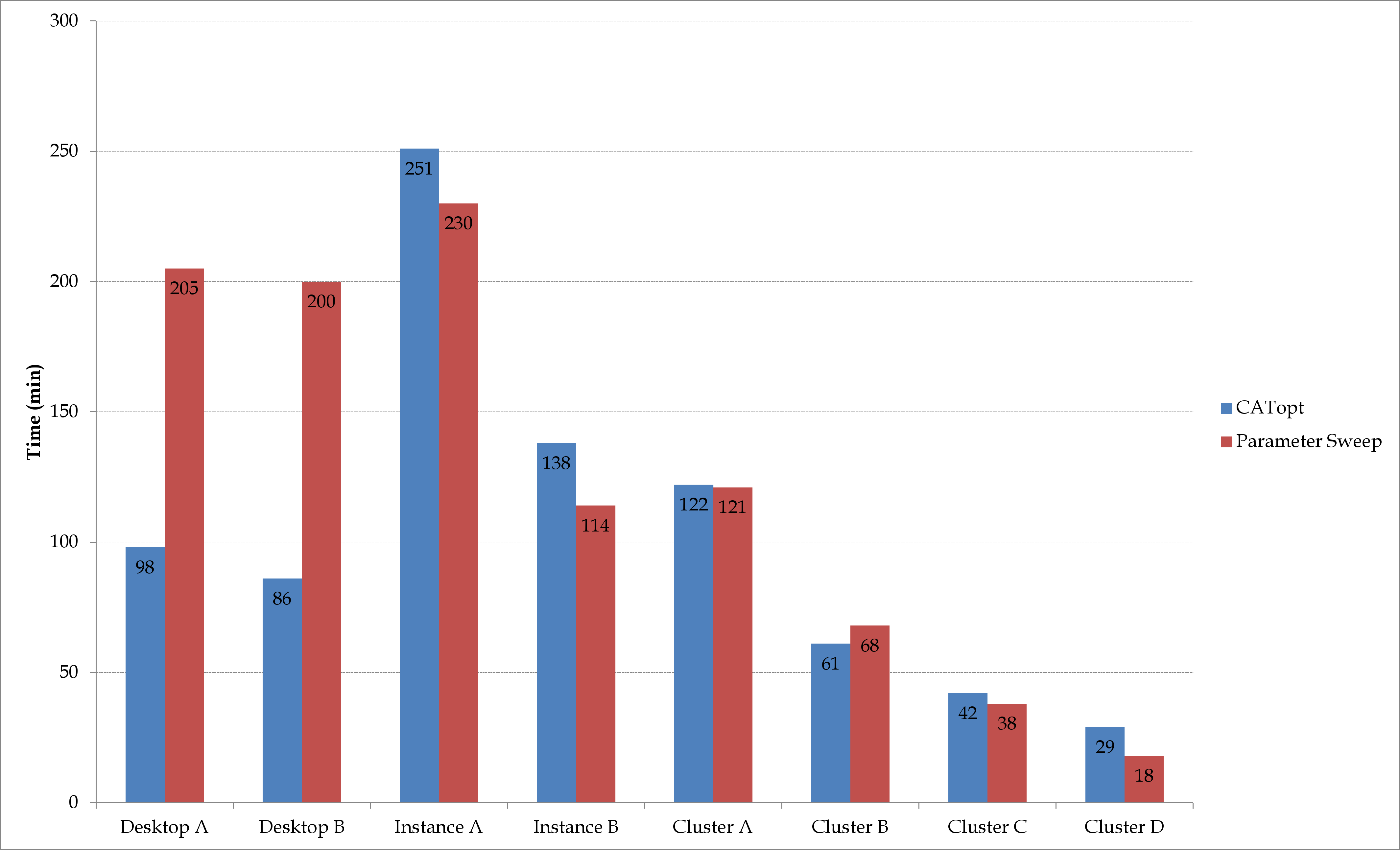}
	\caption{Best-case timing results of the CATopt and Parameter Sweep Problems using P2RAC}
	\label{graph2}
\end{figure}

Figure \ref{graph1} is a graph showing the relative speed-up achieved for both the experimental problems with increasing number of Amazon instances. Figure \ref{graph2} is a bar graph that shows the timing for the best results on the two desktops, the two Amazon instances and the four Amazon clusters. The best performance is achieved on Cluster D.

For the CATopt problem in these experiment, the population size is set to 200 and the maximum generations is set to 50. In both the problems there is a near 100\% efficiency for up to 4 Amazon instances, after which there is a drop in the parallel efficiency. The reduction in parallel efficiency is due to increase in communication overheads between virtualised cloud instances. The acceleration achieved for both the problems is satisfactory considering the low cost of the infrastructure employed and that the problems can be directly deployed from an Analyst's site without any additional tuning.

\begin{figure}
	\centering
	\includegraphics[width = \textwidth]{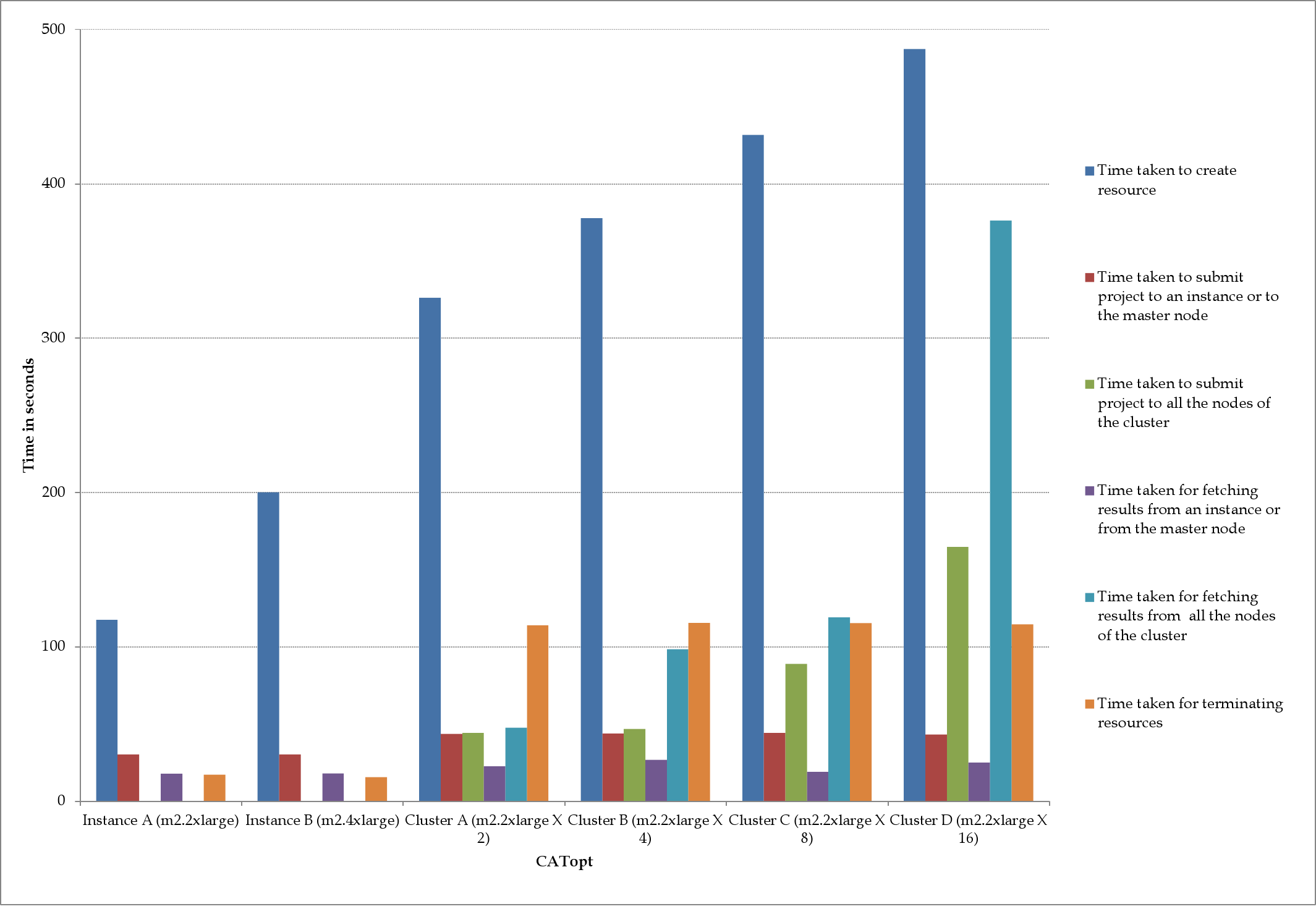}
	\caption{Time taken for creating resource, submitting project, fetching results and terminating resources for the CATopt problem on Amazon instances and clusters}
	\label{graph3}
\end{figure}

\begin{figure}
	\centering
	\includegraphics[width = \textwidth]{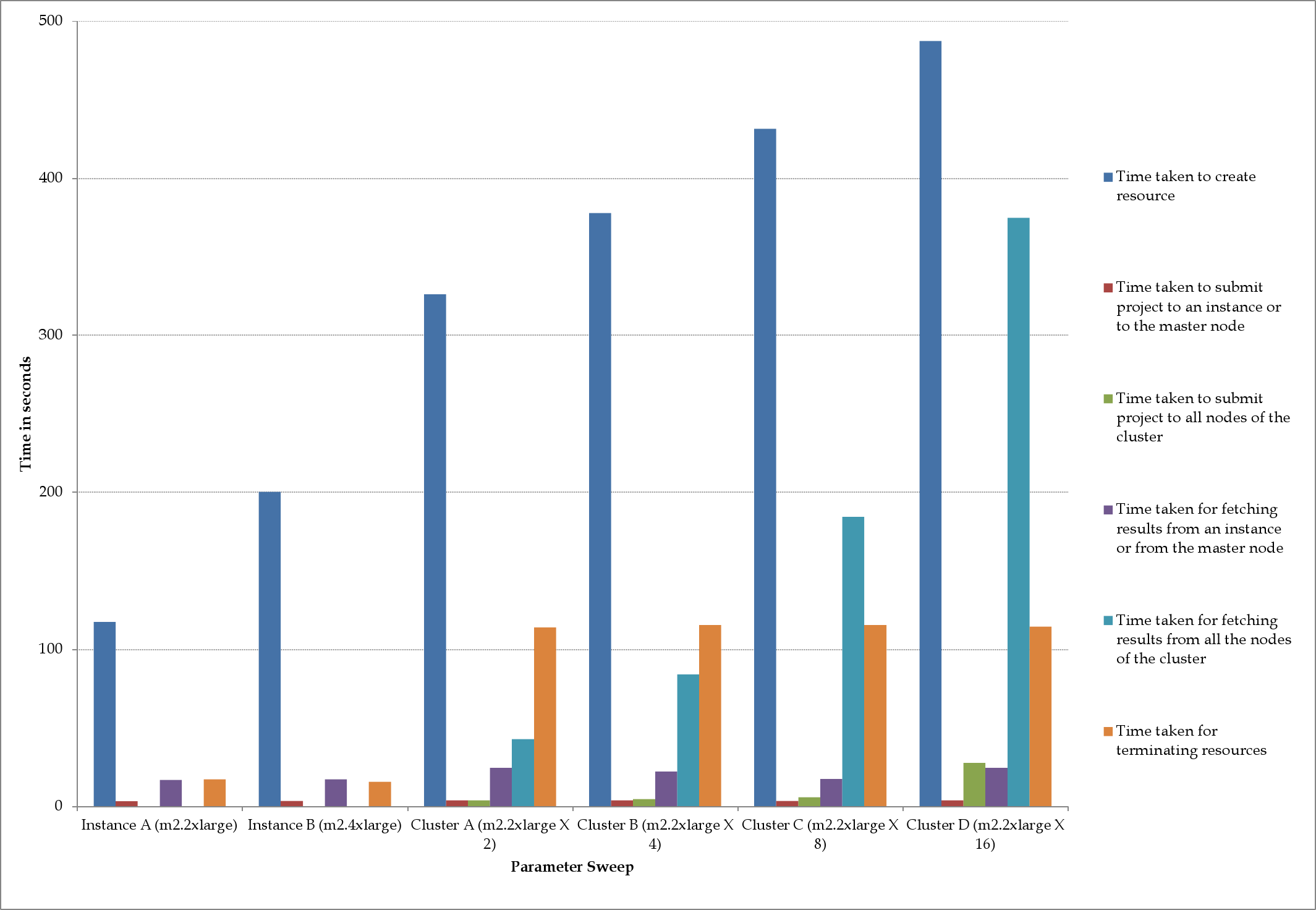}
	\caption{Time taken for creating resource, submitting project, fetching results and terminating resources for the parameter sweep problem on Amazon instances and clusters}
	\label{graph4}
\end{figure}

Figure \ref{graph3} and Figure \ref{graph4} are bar graphs for the CATopt and parameter sweep problems showing (a) the time taken to create the Amazon resource, (b) the time taken to submit the project to an instance or to the master node of a cluster, (c) the time taken to submit the project to all the nodes of a cluster, (d) the time taken for fetching results from an instance or from the master node of a cluster, (e) the time taken for fetching results from all the nodes of a cluster, and (f) the time taken for terminating the resources. 

There is an increase in the time taken for creating Amazon instances and clusters. Though the creation of resources occur in parallel, nearly 7 minutes are required to initialise a 8 node m2.2xlarge cluster and approximately 8 minutes are required to initialise a 16 node m2.2xlarge cluster. This indicates that the time taken for configuring large-scale clusters will also increase. Alternative techniques will need to be investigated to reduce this time and incorporated within the underlying interface that manages AWS. The time taken for terminating the instances and clusters remain the same. 

The time taken to synchronise the project directory comprising the R script and the data (for the first experiment 300 MB and for the second 3 MB) on an instance or on the master node of the cluster and the time taken for fetching results from an instance or from the master node of a cluster remains the same on all resources considered in the experiment. However, there is an increase in both the times for submitting a job and for retrieving results when all the nodes of the cluster are considered. Though the submission of the job and the retrieval of results are parallel in nature there is an increase in time, eliciting the need for investigating other parallel methods that can reduce this time. For large jobs like CATopt that need to run for many hours the additional minutes do not seem significant, however, it may not be worthwhile to spend a lot of time for creating and moving data around resources for small jobs. 

\section{Discussion and Conclusions}
\label{conclusion}

There are a large number of cluster on cloud frameworks supporting a variety of applications. Frameworks such as ElasticR \cite{project1}, CycleCloud \cite{project2} and Pegasus \cite{pegasus} provide a large dashboard for configuring a high-performance cluster on the cloud. These frameworks are useful for developers who have advanced knowledge of the technicalities of the cloud, and can adapt their analytical workloads for such frameworks. Ad hoc analytics will not be easy on such tools as they are mostly performed by Analysts who have limited technical skills, and therefore their prototyping requires a simpler framework. Frameworks such as CloudBLAST \cite{CloudBLAST} and CloudBurst \cite{CloudBurst} support easy workflows without having to specifically adapt analytical workloads for the cloud, but are aimed at computational biological applications and thereby limit their use for Analysts. Frameworks such as Cloudfoundry \cite{project3} and Starcluster \cite{project4} do not provide explicit support for R programming language and therefore a lot of manual procedures are required to run an Analyst's workload. Such frameworks do not provide seamless execution of a task on the cloud. 

In this paper, a platform that (i) provides an Analyst with an interface to seamlessly submit an analytical job and collect its results, and (ii) provides the flexibility to manage the resources and jobs is ideal for an Analyst who needs to exploit the benefits of the cloud. Analytical jobs can benefit from harnessing the benefits of cloud computing such as on-demand availability of resources, scalability of the resources and low costs for maintenance. The Platform for Parallel R-based Analytics on the Cloud (P2RAC) proposed and implemented in this paper provides an interface for an Analyst who needs to submit a R-based analytical job on the cloud. P2RAC is therefore placed in between an Analyst and the cloud infrastructure. P2RAC provides support for instances and clusters on the Amazon cloud and offers a set of core and diagnostic tools which can be used from the command-line. 

The core tools support resource management, data management on the resource and execution management. Resource management functionalities range from gathering resources required for an analytical job and releasing them after their use. Data management ranges from providing the resources with data required for executing an R-script and gathering results from the resources onto the Analyst's site once the R-script has completed execution. Execution management provides functionalities for executing an R-script on the resources acquired.

The diagnostic tools support the testing of resources and provide information of the resources employed by an Analyst. This is facilitated by providing access to log in to the acquired resources.

The feasibility of P2RAC is validated by considering two analytical problems. The algorithm for solving the first problem incorporates co-operative parallelism for optimisation, while in the second problem, multiple independent jobs are employed for parameter sweep. Both the problems are provided from an Analyst work site to the Amazon cloud using P2RAC. The results obtained from P2RAC is an evidence that R-based analytical problems can benefit from cloud computing and P2RAC facilitates the execution of the analyst job on the cloud.

The installation package and the source code for P2RAC is hosted via the Python Package Index (PyPI) http://pypi.python.org/pypi. The installation is built for the Linux OS, supports Virtual Python Environment \cite{VirtualEnvironment} and can be installed on a desktop using Easy Install \cite{EasyInstall}. After installation, P2RAC can be configured automatically using \texttt{ec2configurep2rac}.

In the future, it is anticipated that another version of P2RAC will be released that incorporates the support for spot instances. Fault tolerance through distributed checkpointing for spot instance used in P2RAC will be explored. Further, the dynamic scaling of clusters, i.e., increase and decrease the size of a cluster when required by a job that is executing on the cluster, will be investigated for implementation. While packages such as those reported in \cite{SNOWwebsite, expparallelism1} support explicit parallelism, efforts will be made to provide a dedicated interface that can exploit Amazon EC2 for R programming.

\section*{Acknowledgement}
The authors would like to thank Dr. Georg Hoffman and Dr. Oliver Baltzer of Flagstone Re, Halifax, Canada for their support and participation in this research.

\end{document}